# Ion Irradiation of Nanocrystalline Graphene on Quartz and Sapphire

**Maria Edera and Alexander M. Zaitsev***

College of Staten Island of the City University of New York

2800 Victory Blvd., Engineering Science and Physics, Staten Island 10314, USA

* Corresponding author: alexander.zaitsev@csi.cuny.edu

Tel: 718 982 2812

## ABSTRACT

The effects of $Ga^+$ ion irradiation and high temperature annealing on behavior of nanocrystalline graphene directly grown on quartz and sapphire are presented. It is shown that nanocrystalline graphene stands fairly high doses of ion irradiation (up to $3\times10^{14}$ cm$^{-2}$ of 5 to 50 keV $Ga^+$ ions) without degradation in conductance. At higher doses, nanocrystalline graphene rapidly loses its conductance and at doses over $2\times10^{15}$ cm$^{-2}$ becomes actually insulating. Annealing in vacuum restores conductance of the irradiated nanocrystalline graphene and, if the irradiation has not exceeded a dose of $3\times10^{15}$ cm$^{-2}$, this restoration can be complete. Ion irradiation at high doses approaching $10^{16}$ cm$^{-2}$ results in complete ion sputtering of a few layer graphene. Along with the irradiation-induced reduction of conductance and the temperature-induced restoration of conductance, two other effects of the ion irradiation are the enhancement of adhesion of graphene to substrate and the increase in the nucleation capability of substrate for direct deposition of graphene. For 50 keV $Ga^+$ ions, the enhancement of adhesion is observed at irradiation doses over $2\times10^{14}$ cm$^{-2}$. The promoted graphene nucleation is observed in a broad dose range. It is shown that the above effects can be used for development of methods of patterning of graphene on insulating substrates and a method of imprint lithography of graphene.

**Key words**: graphene, ion irradiation, focused ion beam, graphene annealing, graphene adhesion, graphene sublimation, graphene patterning.

**I. INTRODUCTION**

Graphene, a new prospective electronic nanomaterial, has attracted much attention as an alternative to traditional semiconductors in a number of applications [1-8]. Graphene has many advantageous physical properties making it unique among other electronic materials. One of them is the combination of a fair electrical conductance and very high optical transparency in a broad spectral range. This makes graphene a material of choice for optoelectronics and photovoltaics where high optical transparency is required, but the moderate electrical conductance is not an issue. Another one is the combination of the true 2D atomic structure and a very high mobility of charge carriers. This makes conductance of graphene very responsive to the presence of adsobrates on its surface and hence makes graphene an extremely sensitive electronic chemical sensor [9, 10]. Another advantage of graphene is that it is a true nanoelectronic material. Unlike semiconductors with considerable energy bandgap between the valence and conduction bands, graphene is a semimetal with the zero bandgap energy. Therefore, the unipolar conductivity – the key property of an electronic material - can be induced in graphene without impurity doping but controllably by a low external bias. In contrast, in conventional semiconductors, the unipolar conductivity can be practically achieved only via impurity doping. The impurity doping is one of the major limitations of non-zero bandgap semiconductors, which limits the down-scaling of electronic devices based on the principle of bipolar junction. For instance, silicon field-effect transistor (FET) ceases to work as a practical device at a size of 10 nm [11].

Graphene reveals its highest electronic parameters only when in a form of isolated single crystal layer. However, direct growth of perfect single crystal graphene on surface of standard electronic materials (silicon and silicon oxide) remains an unmet challenge. By now, commercially viable technologies are those based on two-step procedure of CVD growth of graphene on copper followed by transfer of the grown graphene onto the working substrate. The step of the transfer is one of the major limiting factors of these technologies. The recently reported direct CVD growth of graphene on germanium could be a promising alternative [12]. Yet it is not applicable for direct deposition on silicon or $SiO_2$.

A technologically simple and inexpensive alternative to the high quality single crystal graphene is nanocrystalline graphene [13, 14]. Nanocrystalline graphene can be grown by CVD deposition from gas precursor directly on almost any material over large area [15-17]. Although the electronic properties of nanocrystalline graphene are inferior to those of single crystal graphene (low charge carrier mobility in the range 10 to 100 cm$^2$/Vs and rather high growth temperature over 800°C), nanocrystalline graphene possesses reasonably high electrical conductance (in the range 10 to 100 kohm/square), high optical transparency (about 95%) and high chemical sensitivity (even non-structured nanocrystalline graphene may detect the presence of $NO_2$ molecules at a concentration of a few tens of ppb).

When adopting a new material in electronics, along with the technology of its growth, it is equally important to develop a technology of its structuring and patterning. Graphene, as a 2D material, is very suitable for planar patterning, e.g. using lithography. A specific related feature of graphene, which can be used for its patterning, is a very high sensitivity of conductance to radiation damage. It has been shown in a number of publications [18-27], that practically any irradiation (even with electrons of energy as low as 10 keV) damages graphene and reduces its conductance. Upon achieving a critical concentration of defects of 1%, the irradiated graphene becomes actually insulating [28-32]. Although the electron irradiation has been shown to change structural and electronic properties of graphene [33-40], its capabilities are limited. The ion irradiation is a more effective technique in many ways. Firstly, the ion irradiation is much more efficient in the energy transfer from the energetic ions to the graphene atoms and, consequently, in defect production. Secondly, due to the processes of the secondary irradiation by the recoil atoms generated in graphene and in the underlying substrate, the spectrum of the structural defects produced by ion irradiation is much broader than that produced by electrons. Thirdly, the ion irradiation offers unique opportunity for impurity doping (ion implantation). This impurity doping can be achieved both by insertion of atoms from the primary ion beam and via backscattered atoms from the substrate. Fourthly, the ion irradiation is the most precise technique of deterministic 2D and 3D modification of materials at a level down to a few nanometers and the method of addressing of single elements in nanostructures.

The effect of irradiation-induced reduction of electrical conductivity of graphene offers an opportunity of achieving sharp contrast conductor-insulator without physical removal of the graphene layer. If the ion irradiation is performed with focused ion beam (FIB), the transition between the conductive and non-conductive areas can be made as sharp as 1 nm. This opportunity is attracting attention as a way towards the development of a technology of maskless, resist-free patterning and as an approach to fabricate novel graphene-based electronic nanodevices, like polarity-reversible FETs [41] with on-off current ratio by far exceeding that of the conventional graphene FETs [42, 43]. The development of a resist-free technology of patterning could also resolve the issue with the resist residues, which, if left on the surface of graphene, may considerably affect its transport properties [44-48].

Another useful effect of ion irradiation is the improvement of adhesion of thing films to the underlying substrates. The effect of the irradiation-enhanced adhesion has been studied since decades and is well known in material science, chemistry, biology and medicine. Just to give examples we refer to a few references [49-53]. Although the reliable adhesion of graphene to the working substrate is an important technological property, the influence of irradiation and other treatments with energetic particles (e.g. plasma) on adhesion of graphene has not been studied systematically yet [19, 54, 55].

Along with the effects of the irradiation-induced suppression of conductivity and the irradiation-induced adhesion, the irradiation-induced enhancement of nucleation of graphene on substrate is another effect to be considered as beneficial for patterning. By now, this effect is at the very beginning of its studying. The only relevant publication we could find is [56], where the enhanced growth of graphene on SiC was demonstrated using FIB irradiation with $Si^+$ ions.

Below we show that all the effects of ion irradiation on graphene mentioned above can be used for patterning of nanocrystalline graphene as well. Besides, we have found that the restoration of conductance after high temperature annealing and the evaporation at high temperatures are two new effects, which also can be used for patterning of graphene.

## II. EXPERIMENTAL

Nanocrystalline graphene films were grown directly on single crystal quartz and single crystal sapphire substrates from pure methane in an all-graphite furnace according to the CVD method described in [14]. The substrates were purchased from MTI Company. According to the specifications, they were cut perpendicular to z-axis and polished to a finish Ra < 1 nm. Before any processing, all substrates were ultrasonically cleaned in acetone. Parameters of the growth were in the range: methane pressure from 1 to 5 mbar, temperature from 800 to 1400°C, growth time from 0.5 to 30 min. Typical dependences of the growth rate on temperature, pressure and time are shown in Figure 1.

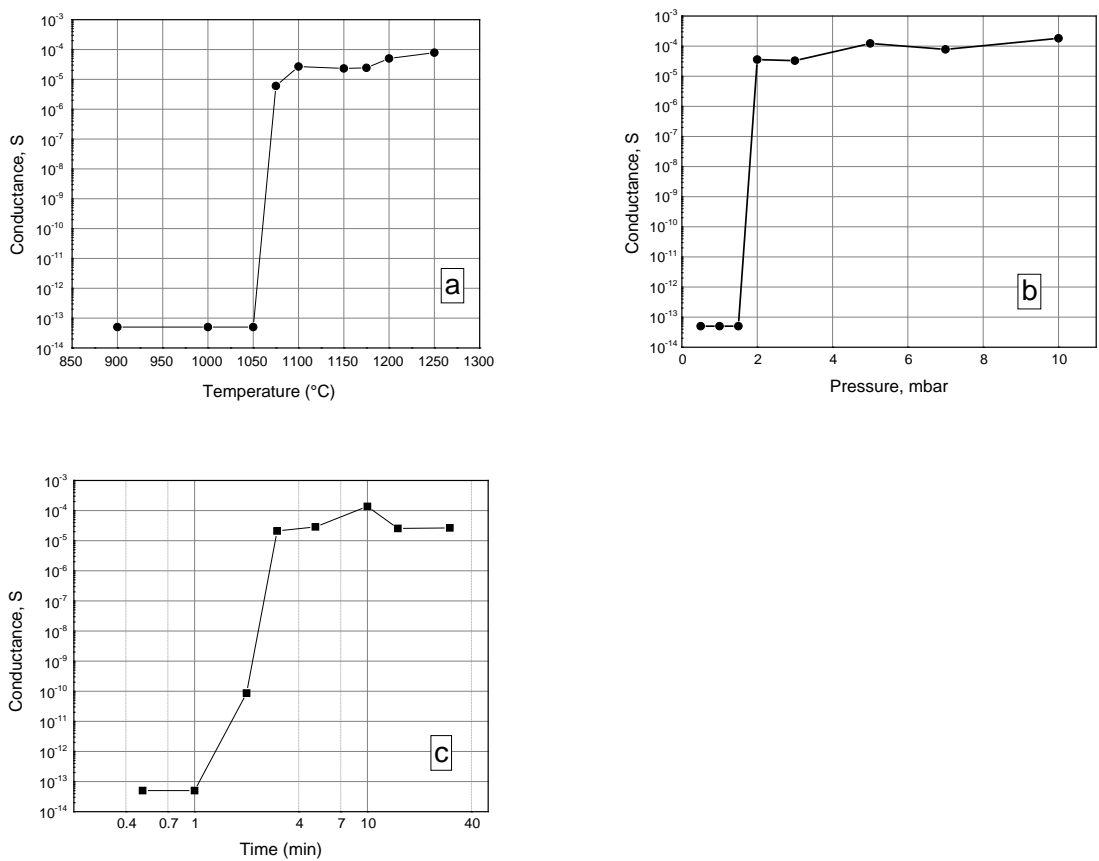

FIG. 1. Conductance of carbon nanofilm grown in different regimes: (a) temperature dependence of growth at a pressure of 5 mbar and time 15 min; (b) pressure dependence of growth at a temperature of 1200°C and time 15 min; (c) time dependence of growth at a temperature of 1200°C and a pressure of 5 mbar. All three dependences exhibit a threshold-like behavior.

There are thresholds of temperature, pressure and time at which continuous conductive film starts to form. The threshold value of each parameter depends on the other parameters as well as on the surface roughness, the surface termination and its atomic structure (presence of defects). Thus the growth temperature, the methane pressure and the growth time could be chosen so that a continuous conductive film formed only on the surfaces of particular physical and chemical parameters. This peculiarity was used for selective growth of graphene on predetermined areas. For this, the growth parameters were set so that no growth occurred on pristine areas of substrate, while reliable growth on ion irradiated areas took place.

The carbon films grown for this research were homogeneous, transparent and possessed conductance at a level of a few tens of kohm/square. As shown in [14], these films can be identified as a few layer nanocrystalline graphene of a thickness of 1 nm.

Electrical conductance of graphene films was measured at room temperature using Keithley 2300 analyzer and tungsten needles placed on the sample surface with identical force. Since we had to perform hundreds of measurements in different places of many samples, we practically could not use a four probe method with deposition of metal electrodes. Instead, a simple two probe method was used. The distance between the needle tips touching graphene was about 20 microns. Most of the measurements were performed at a voltage of 1 V. This voltage was low enough to avoid non-linear effects in the current flow. The level of sensitivity of the measurements was in the range 0.01 to 0.1 pA. The quality of the electrical contacts between the tungsten tips and the graphene layer was not always good and that could cause considerable fluctuations of the current. After having performed many measurements on different samples at different voltages and at different conditions of placing probes on sample surface, we estimated the experimental error of the measurements as an order of magnitude. Thus, the experimental error bar on all the graphs showing conductance is implied to be an order of magnitude. However, even this seemingly high experimental error does not compromise the conclusions made on the basis of the electrical measurements since the changes in conductance occurred over many orders of magnitude.

Ion irradiation was performed with $Ga^+$ ions using focused ion beam (FIB) system Micrion 2500. The ion energy varied from 5 to 50 keV at an ion beam current in the range from 0.6 to 1.3 nA. Multiple square areas of size $150\times150$ $\mu m^2$ were irradiated with doses in a range from $10^{13}$ to $10^{16}$ $cm^{-2}$ by scanning ion beam. In order to ensure homogeneity of irradiation, the ion beam was defocused to a spot of 2 µm. Vacuum in the sample chamber during ion irradiation was better than $2\times10^{-7}$ mbar. Thus we believe that no considerable contamination of the sample surface occurred during irradiation.

After irradiation, the samples were annealed at temperatures from 400 to 1500°C in the same graphite furnace, which was purged with Ar (ultra-high purity grade) and evacuated to a pressure below $10^{-5}$ mbar. The furnace allowed controllable annealing time from 10 seconds to many hours. Since all hot parts of the furnace (sample container, heater, shields and electrodes) were made of high purity graphite, we assume that there was no any noticeable contamination of the sample surface during annealing.

Optical images of the irradiated structures were taken with Nikon Eclipse LV100 microscope in the regimes of transmitted and reflected light. Visualization of the irradiated areas was an issue and this should be addressed separately. It was not problematic to find the irradiated squares when the conductivity contrast between the irradiated and non-irradiated areas was considerable, e.g. for high dose irradiation. A very helpful feature in this respect was a high optical sensitivity of quartz to irradiation. Ion irradiation, even with relatively low doses, makes quartz permanently gray and this color change is seen in microscope. In contrast, sapphire is not this sensitive to ion irradiation. A test irradiation of pristine quartz and sapphire substrates was performed in the regimes used for the irradiation of the samples with graphene. It was found that the quartz substrates changed their color in almost all irradiated areas, whereas the sapphire substrates did not show any detectable changes in color (Figure 2). Thus, the color changes observed on the graphene-on-quartz samples could be caused both by graphene and quartz, while the color changes observed on the graphene-on-sapphire samples were due to graphene only.

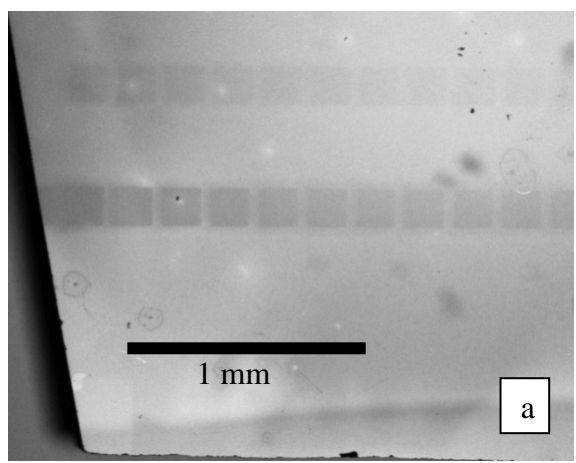
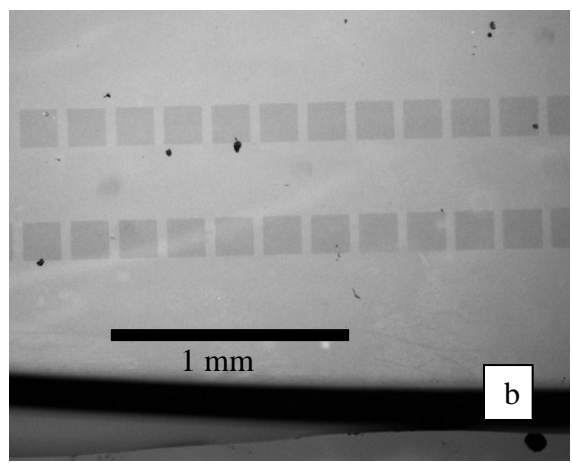
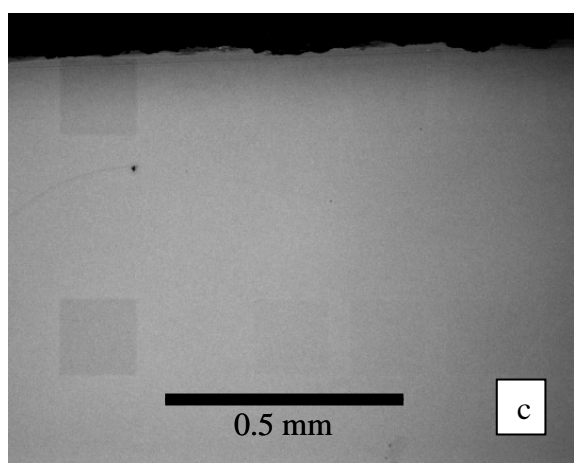
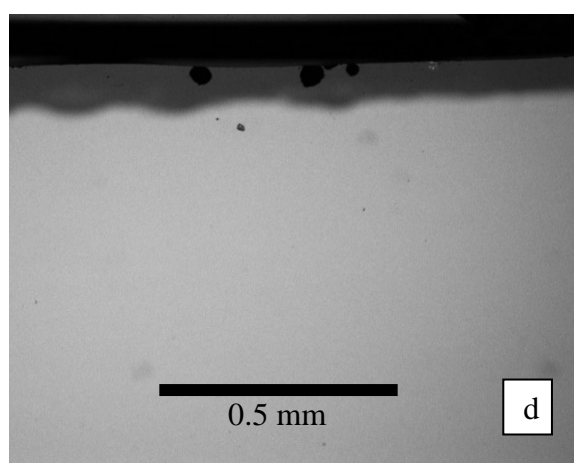

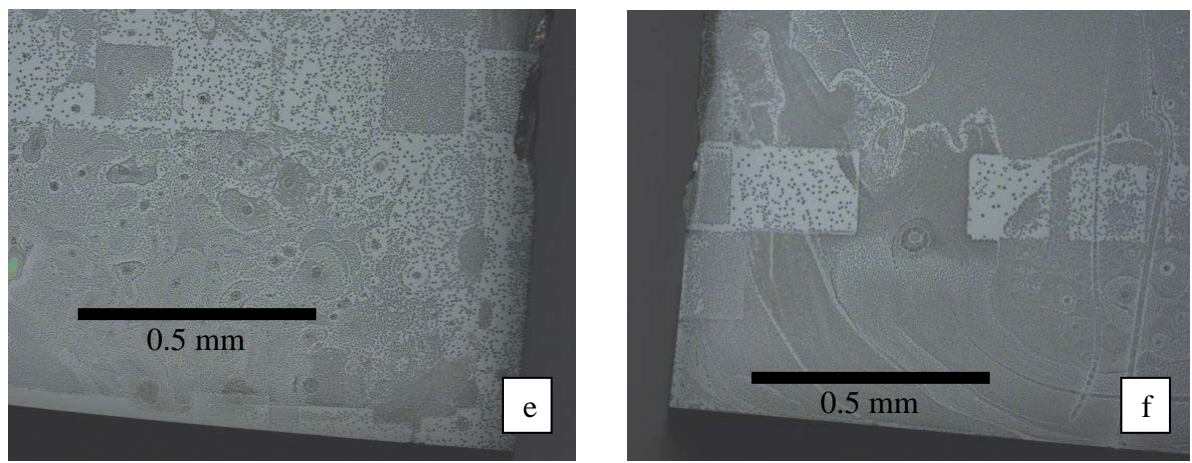

FIG. 2. Optical images in reflected light of ion-irradiated areas on quartz (a, b) and sapphire (c, d) substrates after irradiation (a, c) and after subsequent annealing at 1400°C and complete removal of graphene (b, d). The irradiation dose for different squares varies from $10^{14}$ to $10^{16}$ cm$^{-2}$. All the irradiated squares on quartz are clearly seen immediately after irradiation and after high temperature annealing and removal of graphene. On sapphire, only squares irradiated at high doses (left squares on (c)) are clearly seen. The low dose irradiations shown on picture (c) are hardly seen if at all. After high temperature annealing and complete removal of graphene, all irradiated squares on sapphire are not visible. (e, f) Condensation of water on sapphire substrate covered with graphene after ion irradiation. The pattern of the condensated water droplets allows recognition of all irradiated areas.

The problem with the visualization of the irradiated areas on sapphire could be overcome by gentle blowing warm air saturated with water vapor onto the samples. Tiny water droplets condensate immediately on the sample surface and make the irradiated squares clearly seen (Figures 2e, 2f). This effect of the selective water condensation was observed both on sapphire and quartz and was used for the precise positioning of the measuring probes over the irradiated areas.

## III. RESULTS AND DISCUSSION

### A. Conductance of As-Irradiated Graphene

Dose dependences of the conductance of irradiated samples are shown in Figure 3.

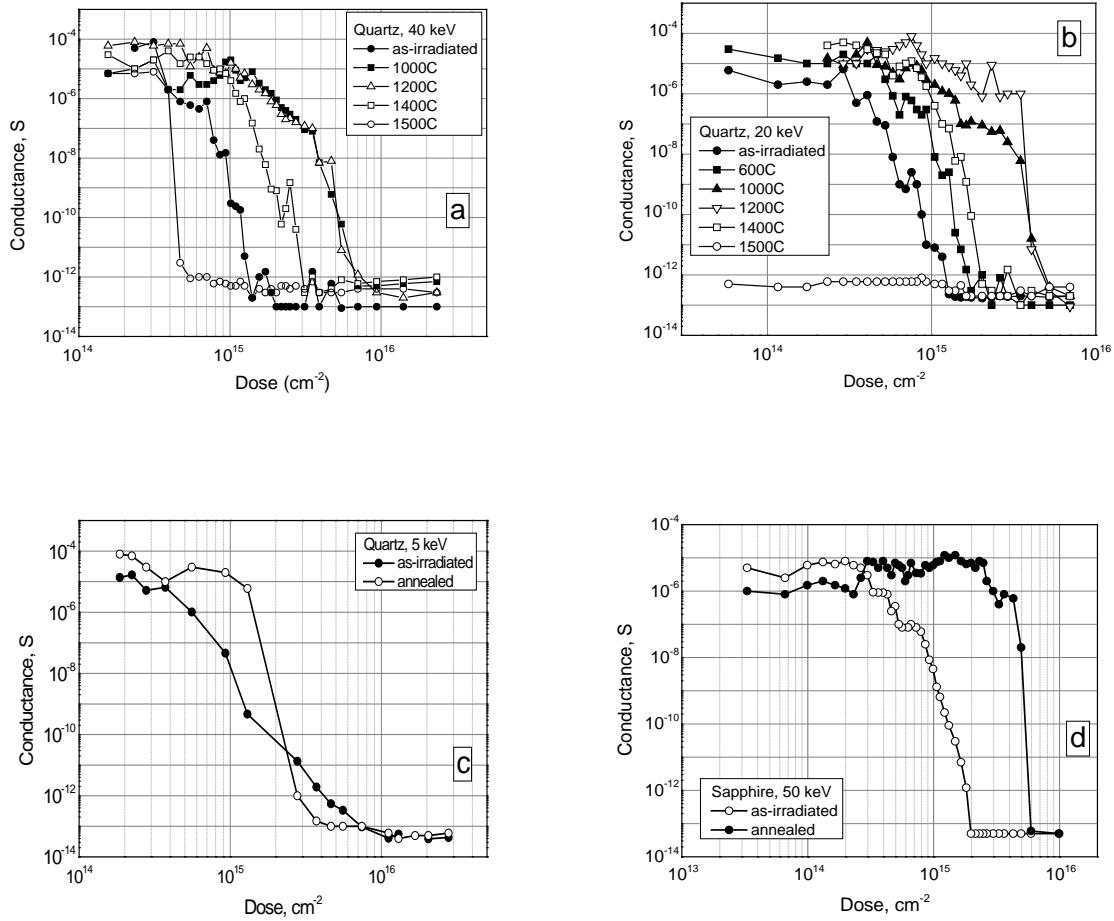

FIG. 3. Dose dependence of conductance of graphene on quartz after irradiation with $Ga^+$ ions of energies 40 keV (a), 20 keV (b), 5 keV (c) and on sapphire after irradiation with 50 keV $Ga^+$ ions (d). Annealing time for all temperatures but 1500°C was 5 to 10 min. The annealing at temperature 1500°C was performed for 1 min.

The ion irradiation with doses below $3\times10^{14}$ cm$^{-2}$ has not affected the conductance of nanocrystalline graphene both on quartz and sapphire substrates. However, a dose of $3\times10^{14}$ cm$^{-2}$ of $Ga^+$ ions is sufficient to considerably suppress the conductance of single crystal graphene. An obvious explanation of this difference is the inherent disorder of nanocrystalline graphene [14]. This critical dose $3\times10^{14}$ cm$^{-2}$ does not depend much on the ion energy (at least in the range of a few tens of keV) and can be taken as a phenomenological parameter of the atomic disorder in nanocrystalline graphene. Based on this dose only we cannot give a meaningful estimation of the concentration of defects. The reason for that is that the ion irradiation damage of graphene on substrate is a very complex process involving the primary damage by fast ions, the damage by recoil atoms from the substrate, the implantation of atoms from the substrate, the sputtering as well as the processes of the secondary defect transformations. We can just say that $Ga^+$ ion irradiation with an energy of a few tens of keV at doses up to $3\times10^{14}$ cm$^{-2}$ does not produce more atomic disorder than it is present in nanocrystalline graphene CVD-grown on quartz and sapphire. This conclusion is supported by the similarity of the Raman spectra of nanocrystalline graphene [13] and single crystal graphene irradiated with $Ga^+$ ions [25, 57, 58]. Both spectra are very much alike and typical of low quality graphene (Figure 4). The main features are D-band (at about 1350 cm$^{-1}$), G-band (at about 1580 cm$^{-1}$) and 2D band (at about 2700 cm$^{-1}$). All these bands are broadened suggesting a considerable disorder of crystal structure.

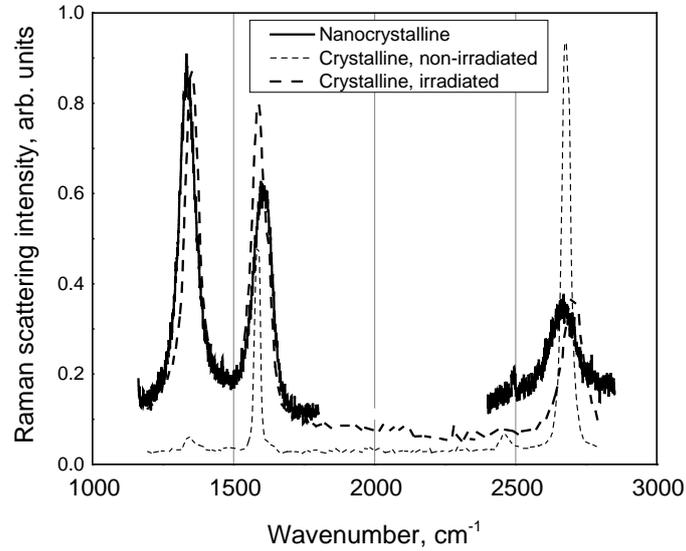

FIG. 4. Raman spectrum of nanocrystalline graphene on quartz (re-plotted from [14]) is compared with that of single crystal graphene irradiated with $Ga^+$ ions to a dose causing an order of magnitude reduction of its conductance (re-plotted from [25]).

At doses over $3\times10^{14}$ cm$^{-2}$, the conductance of irradiated graphene reduces fast and at doses exceeding $2\times10^{15}$ cm$^{-2}$ disappears completely. This irradiation-induced reduction in conductance is an expected behavior for crystalline semiconductors and semimetals, the conductance of which is strongly affected by the defects working as the scattering centers and charge carriers traps [59]. Another reason for the irradiation-induced decrease in conductance of graphene is its gradual ion sputtering and actually its physical removal. Indeed, the dose $2\times10^{15}$ cm$^{-2}$ is comparable with the atomic density of graphene ($3.9\times10^{15}$ cm$^{-2}$). Thus at this dose a considerable fraction of graphene atoms experiences direct collisions with the incident Ga ions and gets knocked out of the graphene layer.

The ion irradiation with the doses of complete suppression of conductance can be used as a method of maskless fabrication of electronic structures on graphene. In this case, the areas with as-deposited graphene are conductive, while the ones with the irradiated graphene are insulating. In Figure 5, the conductance contrast between an irradiated square and the surrounding non-irradiated area is shown. Although a high conductance contrast can be achieved this way, it may disappear after annealing, when the graphene remaining in the irradiated area restores its conductance.

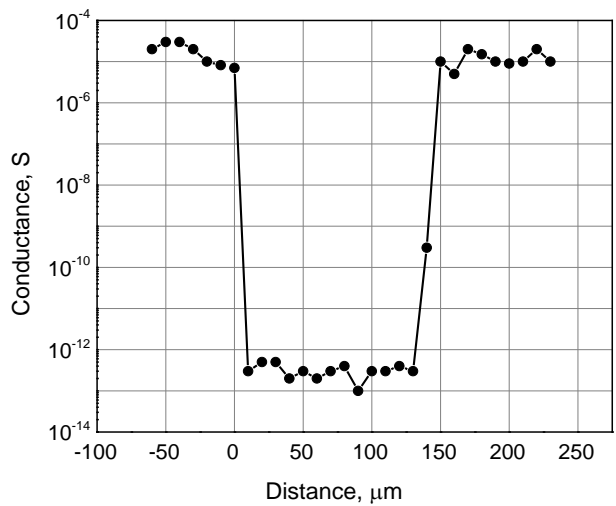

FIG. 5. Change in conductance of nanocrystalline graphene on quartz measured with probes passing across a square as-irradiated with a dose of $5\times10^{15}$ cm$^{-2}$. The irradiated area is completely insulating, while the surrounding non-irradiated graphene exhibits its original conductance of $10^{-5}$ S. Zero mark on horizontal axis is set at one of the edges of the irradiated square.

The observed reduction in conductance of graphene after irradiation is not unique among carbon materials. This effect is also known for bulk graphite. In order to find out whether the ion dose range causing the reduction of conductance is specific of graphene only, we performed comparative measurements on high purity, high density polycrystalline graphite irradiation in the same regimes. It was found that the current-voltage characteristics of the irradiated areas on graphite revealed an electric breakdown behavior (Figure 6). The current-voltage characteristics of this type are typical for conductors covered with thin insulating layer. Thus this result suggests formation of an insulating layer on the surface of graphite after ion irradiation.

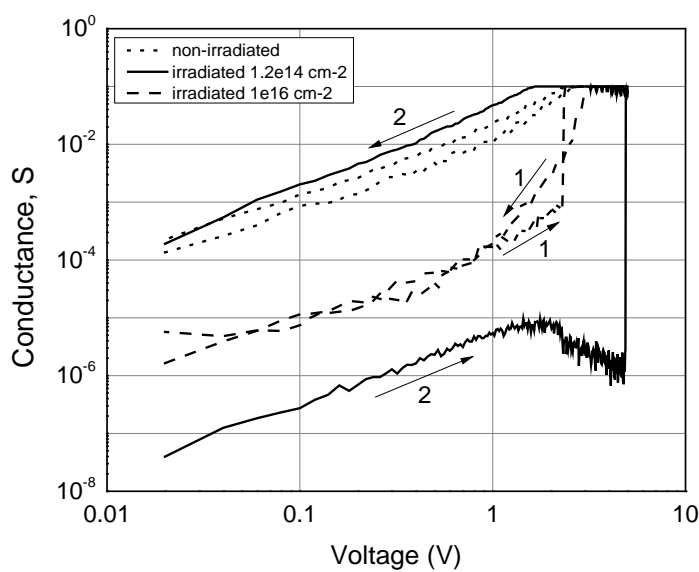

FIG. 6. Current-voltage characteristics measured on the surface of a polished plate of polycrystalline graphite before irradiation (dotted curve) and after irradiation with 50 keV $Ga^+$ ions at doses $1.2 \times 10^{14}$ (solid curve) and $10^{16}$ cm$^{-2}$ (dashed curve). The characteristics of the irradiated graphite exhibit current jump typical for electrical breakdown.

At low voltages, current is proportional to voltage and the conductance remains unusually low till a certain voltage threshold. At this threshold, current jumps up to a high value (compliance limit of the instrument) and, with sweeping voltage down, current follows the dependence of the non-irradiated graphite. This current jump is a typical electrical breakdown through a thing insulating layer formed on the surface. The breakdown starts as an avalanche sustained by injection of high current and then ends up as a thermal breakdown, when the insulating layer gets destroyed. The curve (1) in Fig. 6 is an example of the current-voltage characteristics when the thermal breakdown has not occurred yet and the insulating layer survives. In this case, current returns back to low values when the voltage is reduced. The curve (2) shows current-voltage characteristic of full breakdown with the complete destruction of the insulating layer. In this case, the conductance becomes high at any voltage and riches the value of that of the non-irradiated graphite.

**B. Conductance after Annealing**

After the as-irradiated samples had been measured, they were annealed at different temperatures up to 1500°C. It was found that the annealing at temperature 400°C did not cause noticeable changes in conductance. After annealing at temperatures over 600°C, a considerable recovery of conductance was observed, and, at temperatures from 1000°C to 1300°C, the recovery reached its maximum. The onset of the restoration of conductance at temperatures 600°C to 800°C is to be expected. It is known that vacancies are the most abundant defects in as-irradiated materials. In graphite, vacancies start to move at a temperature of 700°C [60]. Since the atomic structure of a few-layer nanocrystalline graphene is closer to that of bulk polycrystalline graphite than to single crystal single layer graphene, we can also expect similar temperatures of the activation of defect mobility in graphite and nanocrystalline graphene.

At temperatures over 1300°C, a reverse process of reduction of conductance was observed. The most drastic changes occurred at doses in the range from $1 \times 10^{15}$ to $4 \times 10^{15}$ cm$^{-2}$. Figure 7 shows the change in conductance of nanocrystalline graphene on quartz after irradiation at a dose of $10^{15}$ cm$^{-2}$ with Ga$^+$ ions of different energy. It is seen that after annealing at temperatures 1000°C and 1200°C the conductance increases over 7 orders of magnitude and actually restores its original value of $10^{-5}$ S. After annealing at temperatures over 1300°C, the conductance reduces again and completely disappears after annealing at a temperature of 1500°C.

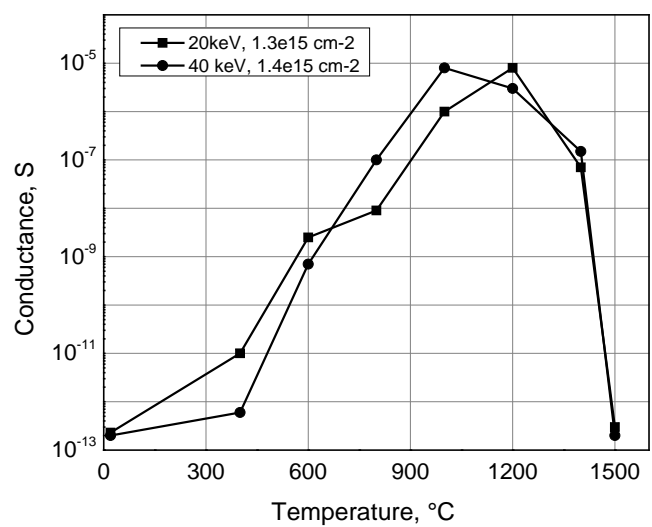

FIG. 7. Conductance of nanocrystalline graphene grown on quartz after irradiation with 20 and 40 keV Ga$^+$ ions at doses $1.3\times10^{15}$ cm$^{-2}$ and $1.4\times10^{15}$ cm$^{-2}$ respectively versus annealing temperature. Although the ion energies differ by a factor of two, there is no essential difference between the annealing dependences.

Complex behavior of conductance of graphene subjected to ion irradiation and annealing suggests involvement of several processes, the obvious ones being the annealing of the irradiation-induced defects and the degradation of graphene at high temperatures. We exclude the surface contamination as a possible cause of the observed conductance change for two reasons: firstly, we took every precaution to minimize the contamination during the irradiation and annealing and, secondly, any feasible surface termination, which could occur in a vacuum with residual pressure below $10^{-5}$ mbar, cannot change the conductance of graphene for many orders of magnitude.

The reason of the degradation of graphene at high temperature could be a strong chemical interaction with substrate and the sublimation of graphene. In order to find out which of these mechanisms prevails, a few samples of graphene on quartz and sapphire were subjected to isothermal annealing at different temperatures in vacuum. Some of these samples were annealed with their graphene-carrying surfaces fully open to vacuum, The graphene-carrying surfaces of the other samples were covered with clean substrates of the same type and same size (stack of two substrates with graphene film in between) and so annealed. The result of this annealing test is shown in Figure 8.

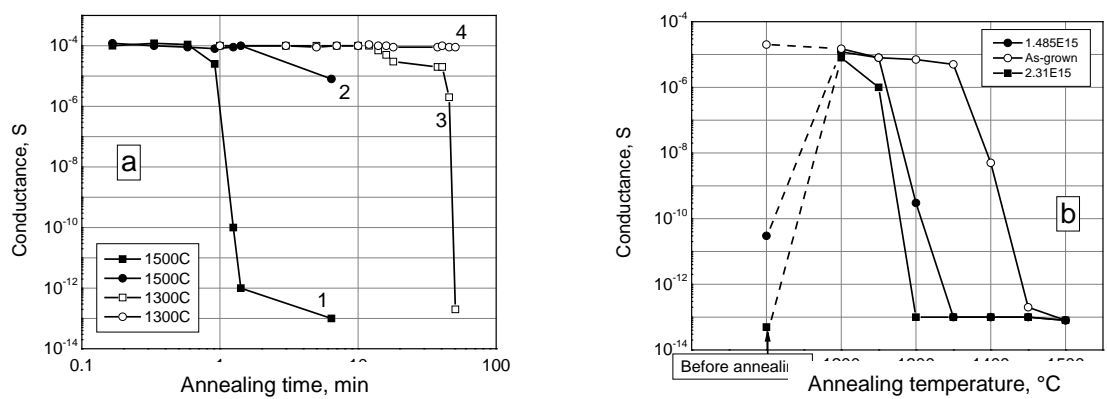

FIG. 8. (a) Change in conductance of non-irradiated graphene film on quartz as a result of isothermal annealing at temperatures 1300°C (3, 4) and 1500°C (1, 2); graphene film open to vacuum (1, 3); graphene film between two quartz plates (2, 4). (b) Change in conductance of graphene film on sapphire with annealing: comparison of sublimation of as-grown graphene and graphene irradiated with doses of partial sputtering.

It is seen that at temperature 1300°C, a slow reduction of conductance of non-covered graphene occurs for 1 hour of annealing and then conductance disappears abruptly. At temperature 1500°C, the conductance vanishes in less than 2 minutes. In contrast, the conductance of the covered graphene reveals no degradation after annealing at 1300°C and only a 10 times reduction after annealing at 1500°C for 7 minutes. This behavior well supports the idea of sublimation of graphene. Assuming that the thickness of highly transparent nanocrystalline graphene film is about 1 nm, the sublimation rate at temperature 1500°C can be estimated as 1 nm per minute. This is unexpectedly high value, which is at least three orders of magnitude greater than the sublimation rate of bulk graphite in vacuum [61-64] and free-standing multilayer graphene [65]. We do not have any solid explanation of this discrepancy yet, however the chemical interaction with the substrate and first of all with its volatile component (oxygen atoms) could be the reason of this highly stimulated sublimation.

Combination of the two effects, the healing of the radiation damage and the sublimation, can well explain the changes in conductance during annealing. At temperatures below 1200°C, sublimation is negligible and the restoration of conductance occurs due to healing of the damaged graphene lattice. Full restoration occurs only if the irradiation dose has not exceeded $2\times10^{15}$ cm$^{-2}$. For higher doses, from $2\times10^{15}$ to $6\times10^{15}$ cm$^{-2}$, conductance restores only partially. For doses over $6\times10^{15}$ cm$^{-2}$, the irradiated areas do not show any conductance neither after irradiation, nor after subsequent annealing at any temperature. This dose $6\times10^{15}$ cm$^{-2}$ of the total irreversible destruction of conductance is most probably the dose of the complete ion-sputtering of the graphene layer. The dose range from $2\times10^{15}$ to $6\times10^{15}$ cm$^{-2}$ corresponds to a partial sputtering of the graphene film. TRIM simulation [66] of the radiation damage of 1 nm thick carbon film on quartz and sapphire predicts the sputtering yield of carbon atoms by Ga$^+$ ions of energy from 10 to 50 keV to be from 1.6 to 2. Thus the area density of the carbon atoms sputtered by a dose of $6\times10^{15}$ cm$^{-2}$ is expected to be about $1.2\times10^{16}$ cm$^{-2}$. This number is very close to 2D atomic density of a three-layer, 1 nm thick graphene ($1.15\times10^{16}$ cm$^{-2}$). Experimental studies of the ion beam sputtering of bulk graphite with 5 keV Ar$^+$ ions give a value of sputtering yield of 1.5 [67]. This number is in good agreement with our data. Thus we may conclude that the total destruction of conductance of graphene after high dose ion irradiation is the result of the physical sputtering of the graphene film and that the ion sputtering is one of the major effects causing the degradation of graphene during ion irradiation.

It is seen in Fig. 2 that the conductance of the ion-irradiated graphene increases with annealing temperature up to 1200°C and then disappears after annealing at higher temperatures due to sublimation. The conductance vanishes first in the areas irradiated with high doses and then all irradiated areas become non-conductive. This behavior is quite expected as the sublimation efficiency of materials strongly depends on their structural quality. The higher sublimation rate of the irradiated graphene is an effect, which can be used for patterning. The regimes of irradiation and annealing can be found at which the irradiated graphene is completely removed, whereas the non-irradiated graphene still retains continuity and a considerable conductance.

**C. Ion Irradiation Enhanced Adhesion**

It has been found that even low dose ion irradiation improves adhesion of graphene grown on quartz and sapphire. The enhancement of adhesion is especially pronounced for quartz substrates. Our tentative explanation of the different efficiency of quartz and sapphire is the different chemical composition and, first of all, the presence of silicon atoms. TRIM simulation of the Ga ion irradiation of graphene on quartz shows that a considerable intermixing of C, Si and O atoms occurs at the graphene-quartz interface. Thus the formation of stable Si-C bonds between graphene and quartz during ion irradiation, and especially after subsequent annealing, is quite possible.

The irradiation enhanced adhesion was found when the samples with the irradiated graphene were gently rubbed with cotton soaked in acetone (Figure 9). Usually rubbing with cotton could easily remove as-deposited graphene from all substrates. In contrast, the ion-irradiated graphene could not be easily removed this way and, if on quartz, it could not be removed completely even after vigorous rubbing.

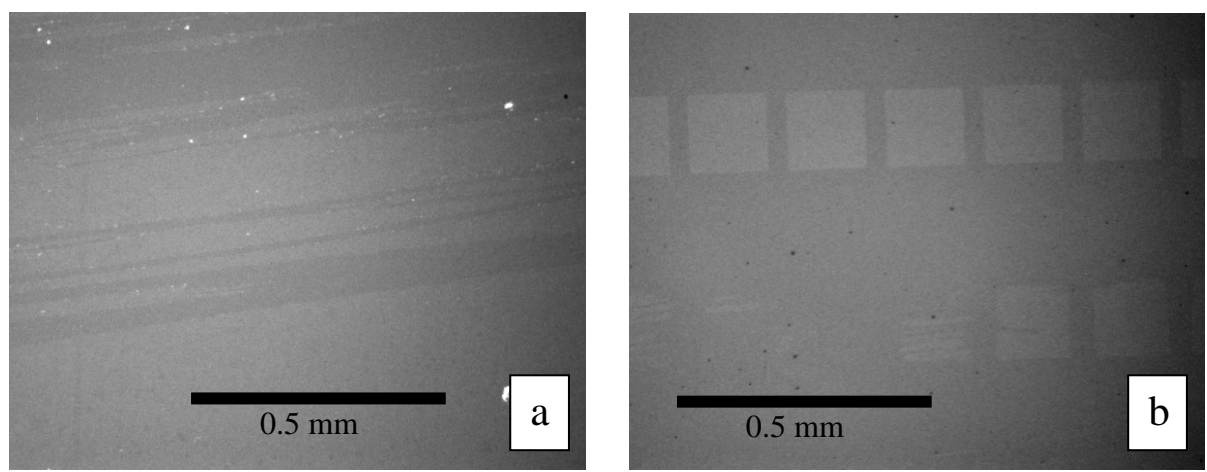

FIG. 9. (a) As-grown graphene of quartz substrate. Considerable removal of graphene film occurs after single gentle stroke with cotton soaked in acetone. (b) Graphene on sapphire substrate after irradiation and annealing at 1200°C. Graphene film remains on irradiated square areas after multiple strokes with cotton soaked in acetone, while it is completely removed from the non-irradiated areas. Graphene uniformly covers the areas irradiated with doses over $10^{14}$ cm$^{-2}$ (upper row of squares), whereas it has been partially removed from the areas irradiated with doses below $10^{14}$ cm$^{-2}$ (bottom raw of squares).

For 50 keV $Ga^+$ ions, the irradiation-stimulated adhesion of graphene occurs at doses over $2\times10^{14}$ $cm^{-2}$ and its strength comes to maximum at doses over $7\times10^{14}$ $cm^{-2}$ (Fig. 10a). The dose $7\times10^{14}$ $cm^{-2}$ is low enough not to affect much the conductance of nanocrystalline graphene (Figure 3).

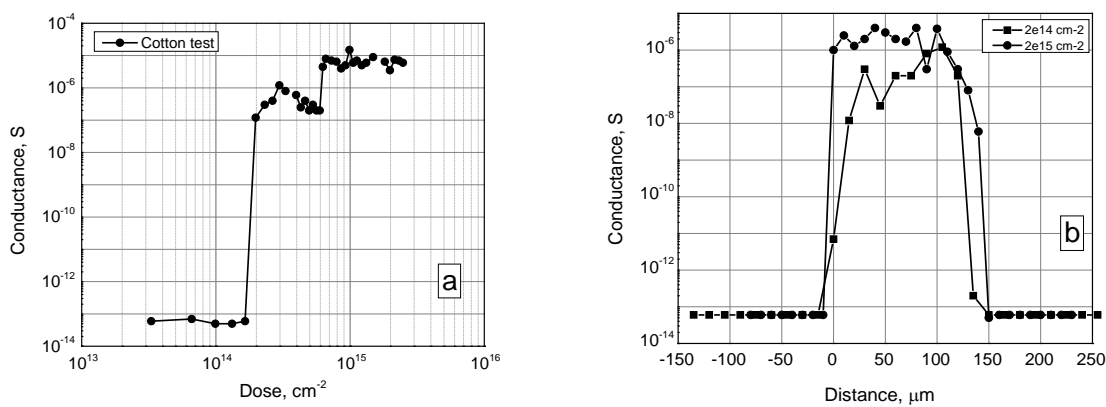

FIG. 10. Conductance of graphene on sapphire after 50 keV $Ga^+$ ion irradiation, annealing at a temperature of 1300°C and single stroke with cotton over the sample surface. (a) Dose dependence of the irradiation-enhanced adhesion. At doses below $2\times10^{14}$ $cm^{-2}$, the adhesion remains negligible and graphene is wiped away completely. In the dose range from $2\times10^{14}$ $cm^{-2}$ to $7\times10^{14}$ $cm^{-2}$, the adhesion improves and graphene remains partially on the irradiated areas. At doses over $7\times10^{14}$ $cm^{-2}$, complete retention of the graphene layer takes place. (b) Conductance contrast between the irradiated and non-irradiated areas. Conductance is measured with the probes scanning over two areas irradiated with doses $2\times10^{14}$ $cm^{-2}$ and $2\times10^{15}$ $cm^{-2}$. Graphene is completely removed from the non-irradiated surface leaving it nonconductive. The irradiated areas retain conductance due to remaining graphene film. The zero mark on horizontal axis is set at one of the edges of the irradiated area.

The enhancement of adhesion takes place immediately after the irradiation. Subsequent annealing results in a further improvement of the adhesion. After the irradiation and annealing, the adhesion of graphene on quartz may become so strong that the conductive graphene film cannot be removed from the substrate even by an intense rubbing.

The effect of the irradiation-enhanced adhesion of graphene to the substrates can be used for the patterning and development of imprint lithography for graphene. We have found that the application of a sticky tape to the samples with the irradiated graphene and then its peeling off is a way to remove the non-irradiated graphene and to leave the irradiated graphene areas in place. When scanning the measuring probes over the irradiated areas of these samples, a high conductance contrast between the irradiated areas with graphene and the surrounding non-irradiated areas has been revealed. This contrast was similar to that shown in Figure 10b.

**D. Deposition of Graphene on Ion-Irradiated Substrates**

Ion irradiation can be also used for selective growth of nanocrystalline graphene on quartz and sapphire. It has been found that the formation of continuous well conductive graphene on ion-irradiated areas can be achieved in the temperature-pressure-time regimes, at which no conductive film grows on non-irradiated surface (Figure 11).

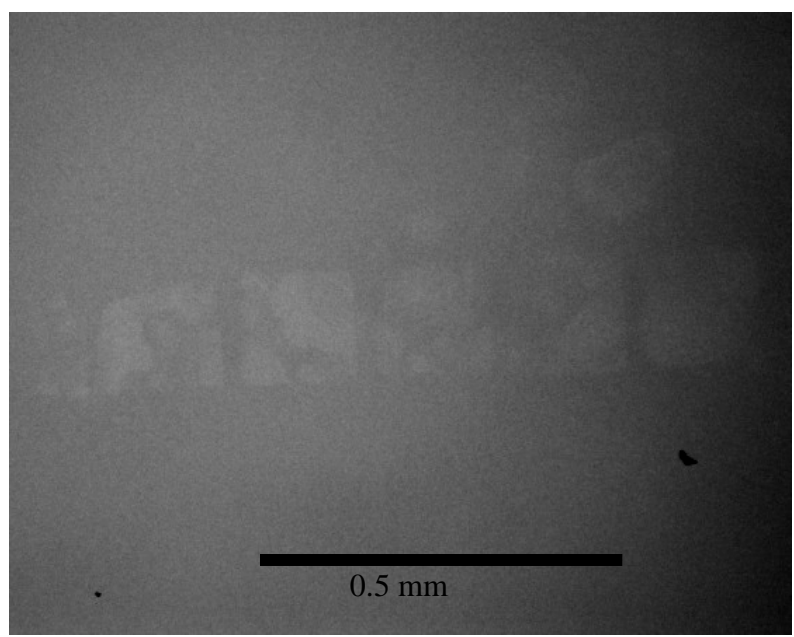

FIG. 11. Deposition of graphene on sapphire substrate at temperature 1200°C, methane pressure 1 mbar, growth time 1 minute. At these parameters, visible and conductive graphene film grows only on the ion-irradiated areas.

Our experiments show that the higher the pressure of methane the broader the dose range of the stimulated growth. For instance, at a pressure of 0.1 mbar and temperature 1280°C, the growth of graphene is observed on the areas irradiated with doses from $7\times10^{14}$ cm$^{-2}$ to $2\times10^{15}$ cm$^{-2}$, whereas at a pressure of 2 mbar and the same temperature all the areas irradiated with doses from $2\times10^{14}$ cm$^{-2}$ to $6\times10^{15}$ cm$^{-2}$ reveal deposition of highly conductive film (Figure 12). In both cases, the non-irradiated surface remains insulating.

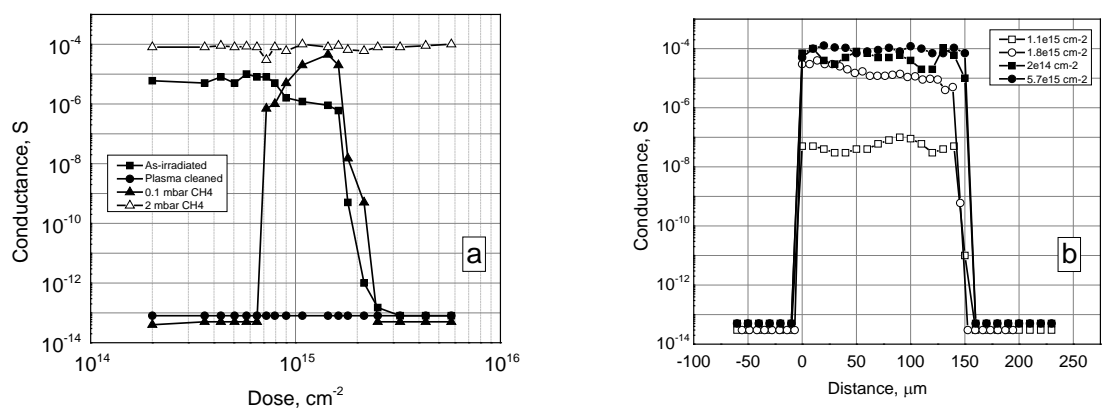

FIG. 12. Deposition of nanocrystalline graphene on quartz ion-irradiated with different doses. (a) Conductance of the irradiated areas after: irradiation and deposition (solid squares); cleaning in oxygen plasma (solid circles); plasma cleaning and deposition at methane pressure of 0.1 mbar (solid triangles); plasma cleaning and deposition at methane pressure of 2 mbar (open triangles). (b) Measuring conductance with probes moved over the squares after irradiation with doses $1.1\times10^{15}$ and $1.8\times10^{15}$ and graphene deposition at 0.1 mbar pressure (open symbols); after irradiation with doses $2\times10^{14}$ and $5.7\times10^{15}$ cm$^{-2}$ and graphene deposition at 2 mbar pressure (solid symbols).

The enhancement of nucleation of graphene on ion-irradiated quartz and sapphire is a very stable effect. It persists after multiple cleanings in solvents, high temperature annealing and even after cleaning in oxygen plasma. This stability suggests that the enhanced nucleation is the result of permanent ion damage of the substrate and, as such, remains until the ion-damaged layer is removed to its whole depth.

## IV. CONCLUSION

We show that using ion irradiation it is possible to control the electrical conductance of nanocrystalline graphene on quartz and sapphire over many orders of magnitude. It is shown that the nanocrystalline graphene stands fairly high doses of ion irradiation (up to $3\times10^{14}$ cm$^{-2}$ for 5 to 50 keV Ga$^+$ ions) without degradation in conductance. At higher doses, nanocrystalline graphene rapidly loses its conductance and at doses over $2\times10^{15}$ cm$^{-2}$ becomes actually insulating. Annealing in vacuum at temperatures over 600°C restores the conductance of the ion-irradiated nanocrystalline graphene and, if the irradiation does not exceed a dose of $3\times10^{15}$ cm$^{-2}$, this restoration can be almost complete. This reverse increase in conductance after annealing occurs via the healing of the radiation damage. The ion irradiation at very high doses approaching $10^{16}$ cm$^{-2}$ results in the complete sputtering of a few layer graphene. It has been also found that along with the radiation damage and the ion beam sputtering, the high temperature sublimation is an important effect involved in the reduction of conductance of the ion-irradiated and annealed nanocrystalline graphene.

It is also shown that ion irradiation improves adhesion of graphene to quartz and sapphire substrates and increases the efficiency of the graphene nucleation on these materials during CVD growth. For 50 keV Ga$^+$ ions, the irradiation-enhanced adhesion is observed at doses over $2\times10^{14}$ cm$^{-2}$. The promoted graphene nucleation is observed in a broad dose range.

It is shown that the above effects can be used for the development of methods of patterning of graphene deposited on insulating substrates and the development of methods of imprint lithography of graphene. As the modern FIB systems allow ion irradiation with a sub 10 nm resolution [24, 68, 69], the patterning of graphene at a few nanometer scale is feasible. It is important that the patterning done this way is maskless and resist-free, and, as such, would not require the steps of cleaning of graphene surface from the resist residues.

**Acknowledgements**

This work was supported, in part, by PSC CUNY Research Foundation, Grant # 67084-00-45. The authors acknowledge the help of Mr. Valery Ray, PBS&T, MEO Engineering Co., Inc., in the maintenance of the FIB system and technical assistance. The authors thank Dr. Searhei Samsonau for valuable advices on CVD growth of graphene and useful discussions.